\documentclass [12pt,a4paper,reqno]{amsart}
\usepackage {graphicx}
\usepackage {longtable}
\usepackage{anysize}
\marginsize{2.5cm}{2.5cm}{2.5cm}{2.5cm}

\begin{document}
\title{On the behavior of solution of nonlinear equations}
\maketitle
\begin{center}
\textbf{Tair  Gadjiev, Sardar  Aliev, Rafig Rasulov }
\end{center}

\begin{center}
\textbf{(Institute Mathematics and Mechanics, Baku State
University)}
\end{center}
\begin{abstract}
In this paper we establish of the Wiener criterion for solution
the mixed boundary problem for nonlinear elliptic equation of
second order.
\end{abstract}
\bigskip

\section{ Introduction and preliminaries.}
Let us consider the problem

\begin{equation}
\label{eq1}
A\left( {u} \right) = \frac{{d}}{{dx}}a_{i} \left( {x,u,u_{x}}  \right) +
a\left( {x,u,u_{x}}  \right) = 0
\end{equation}

\begin{equation}
\label{eq2}
\left. {u} \right|_{\Gamma _{1}}  = 0,\,\,\,\,\,a_{i} \left( {x,u,u_{x}}
\right)\left. {cos\left( {n,x} \right)} \right|_{\Gamma _{2}}  = 0
\end{equation}

\noindent in the domain $\Omega$. Let $\Omega$ be an open set in
$R^{n}$ with the boundary $\partial \Omega = \Gamma _{1} \cup
\Gamma _{2}$, and let $\left\{ \begin{array}{l}
 A(u) = 0 \\
 u - f \in W_{m,0}^1 (\Omega ) \\
 \end{array} \right.$
 be a fixed number. The Dirichlet conditions are
fulfilled in $\Gamma _{1}$, and Neumaun conditions are fulfilled
in $\Gamma _{2}$, and $0 \in \overline {\Gamma _{1}}  \, \cap
\Gamma _{2} $. Moreover we suppose that domain $\Omega $
satisfying isoperimetric conditions. Assume that the functions
$a\,_{i} \left( {x,\,\,u,\,\,p} \right)$, $a\,\left( {x,u,p}
\right)$ are defined for $x \in \overline {\Omega}  $ and
arbitrary u,p, are measurable and satisfy the following conditions

\begin{equation}
\label{eq3}
\begin{array}{l}
 a_{i} \,\left( {x,u,p} \right)p_{i} \ge v\left| {p} \right|^{m} - d\left|
{u} \right|^{m} - g \\
 \left| {a_{i} \,\left( {x,u,p} \right)} \right| \le v\left| {p} \right|^{m
- 1} - b\left| {u} \right|^{m - 1} + l \\
 \left| {a_{i} \,\left( {x,u,p} \right)} \right| \le \,ñ\left| {p}
\right|^{m - 1} + d\left| {u} \right|^{m - 1} + f \\
\sum\limits_{i = 1}^{n} {\left[ {a_{i} \,\left( {x,u,p} \right) -
a_{i} \,\left( {x,u,q\left. {} \right)} \right]\,\,} \right.}
\left( {p_{i} - q_{i}}  \right) \ge c\left| {p - q} \right|^{m}\\
 \end{array}
\end{equation}

\noindent
Here $v,d,g,b,l,c,f-$ const's.

The function $u\left( {x} \right) \in W_{m,0}^{1} \left( {\Omega}  \right)$
is said to be a generalized solution of problem (\ref{eq1}),(\ref{eq2}) if it satisfies the
following integral identity

\begin{equation}
\label{eq4}
\int\limits_{\Omega}  {\left[ {a_{i} \left( {x,u,u_{x}}  \right)\varphi
_{x_{i}}  + a\left( {x,u,u_{x}}  \right)\varphi}  \right]} \,dx = 0
\end{equation}

\noindent for $\forall \varphi \in W_{m,0}^{1} \left( {\Omega}
\right)$. Here $W_{m,0}^{1} \left( {\Omega}  \right)$ is a closure
in $W_{m}^{1} \left( {\Omega}  \right)$ of functions from $
C_0^\infty  (\partial \Omega \backslash \Gamma _2)$.

The principal model operator is the p-laplacian

\[
 - \Delta _{m} u = - div\,\,\left( {\left| {\nabla u} \right|^{m - 2}\nabla
u} \right)
\]

A boundary point $x_{0} $ of bounded $\Omega $ is regular if the solution
$u$to the mixed boundary problem

\[
\left\{ \begin{array}{l}
 A(u) = 0 \\
 u - f \in W_{m,0}^1 (\Omega ) \\
 \end{array} \right.
\]

\noindent has the limit value $f\left( {x_{0}}  \right)$ at
whenever $f \in \,W_{m,0}^{1} \left( {\Omega}  \right)$ is
continuous in the closure of $\Omega $.

In [1]  Wiener proved that in the case of the Laplacian the
regularity of a boundary point $x_{0} \in \partial \Omega $ can be
characterized by a so called Wiener test. In [2]
 Littman, Stampacchia and Weinberger showed that the same Wiener test
identifies the regular boundary points whenever $A$ is a uniformly elliptic
linear operator with bounded measurable coefficients.

For general nonlinear operators the classical Wiener test has to
be modified so that the type $m$ of the operator $A$ is involved.
In [3]
 Maz'ya established that the boundary point $x_{0} $ is regular if
$ W_m (R^n \backslash \Omega ,x_0 )= + \infty$, where $ W_m (R^n
\backslash \Omega ,x_0 ) $ is a Wiener type integral. Later in [4]
 Gariepy and Ziemer extended this result to a very general class of
equation.

In [5]
 Skrypnik established necessary condition of regularity of a boundary points
for general class of equations. However this is necessary condition
coincidenced with a sufficient conditions only in case $m = 2$.

The question whether regular boundary point of $\Omega $ can be
characterized by using the Wiener test has been a well known open
problem in nonlinear potential theory [6]. In case the Dirichlet
condition the problem was partly solved in the affirmative when
[7] proved that if $m$ equals $n$. At last in [8]
 the established the necessity part of the Wiener test for all $m \in \left(
{1,n} \right]$ in case the Dirichlet condition.

In case mixed boundary condition we in [9]
 established a sufficient and a necessary condition of regularity of the
boundary points to a very general class of equations. However this
is necessary condition coincidenced with a sufficient conditions
only in case $m = 2,\,\,m = n,\,\,$or $\,m > n - 1$.
Unfortunately, their method cannot be extended to cover all values
$\,1 < m \le n$.

In this paper we establish the necessity part of the Wiener test for all $m
\in \left( {1,n} \right]$ and prove:

\textbf{Theorem1.1.} \textbf{\textit{Let} }$\Omega $\textbf{\textit{ satisfy
isoperimetric conditions.} }$A$\textbf{\textit{} } \textbf{\textit{finite
boundary point} }$x_{0} \in \overline {\Gamma _{1}}  \cap \Gamma _{2}
$\textbf{\textit{ is regular if and only if}}

\[
W_{m} \left( {B_{1} \left( {x} \right)\backslash \Omega ,x_{0}}  \right) =
\int\limits_{0}^{{{1} \mathord{\left/ {\vphantom {{1} {2}}} \right.
\kern-\nulldelimiterspace} {2}}} {\left[ {{{C_{m} \left( {\Gamma _{1} ,B_{i}
\left( {x_{0}}  \right)} \mathord{\left/ {\vphantom {{C_{m} \left( {\Gamma
_{1} ,B_{i} \left( {x_{0}}  \right)} {\Omega ,\Gamma _{2}}  \right)t^{m -
n}}}} \right. \kern-\nulldelimiterspace} {\Omega ,\Gamma _{2}}  \right)t^{m
- n}}}} \right]} ^{\frac{{1}}{{m - 1}}}\frac{{dt}}{{t}} = \infty .
\]

An immediate corollary is:

\textbf{Corollary1.1.} \textbf{\textit{The regularity depends only on
}}$n$\textbf{\textit{ and}}$m$\textbf{\textit{, not on the operator
}}$A$\textbf{\textit{ itself.}}

\bigskip

Note that no boundedness assumption on $\Omega $ was made in the
theorem above, for we extend the definition of regularity for
boundary points of unbounded sets below. Also observe that the
similar question could be asked also for $m > n$. However, then
all points are regular and the corresponding Wiener integral
always diverges because singletons are of positive
$m$-conductuvity.

The uniformly elliptic linear equations are included in our
presentation. Let us also point out that this methods can be
applied to the equations with weights so that the results of this
paper are easily generalized to cover the equations considered in
[10]. .

Let us give definition of $m$-conductivity. Denote by $F$ bounded
subsets of open set $\Omega $ closed in $\Omega $, and by $G$
bounded open subsets of $\Omega $.

The set $K = {{G} \mathord{\left/ {\vphantom {{G} {F}}} \right.
\kern-\nulldelimiterspace} {F}}$ is called a conductor. By
$V_{\Omega} \left( {K} \right)$ we will denote the class of
functions $\left\{ {f \in C^{\infty} \left( {\Omega}
\right)\,,\,\,f\left( {x} \right) = 1} \right.$, when $x \in F$,
and $f\left( {x} \right) = 0$ when $\left. {x \in {{\Omega}
\mathord{\left/ {\vphantom {{\Omega}  {G}}} \right.
\kern-\nulldelimiterspace} {G}}} \right\}$.

The following quantity will be called a $m$ -conductivity of the
conductor $K$

\[
C_{m} \left( {K} \right) \equiv C_{m} \left( {F,\Omega ,G} \right) =
inf\left\{ {\int {\left| {\nabla f} \right|^{m}dx\,\,:\,\,\,\,\,f \in
V_{\Omega}  \left( {K} \right)\,\,}}  \right\}
\]

Let us formulate conditions for domain. Let $v_{M,m} \left( {t}
\right)$ be the greatest lower bound of $C_{m} \left( {K} \right)$
in the set of all the conductors $k = {{G} \mathord{\left/
{\vphantom {{G} {F}}} \right. \kern-\nulldelimiterspace} {F}}$,
satisfying the condition $m_{n} \left( {F} \right) \ge t,m_{n}
\left( {G} \right) \le M$, where $m_{n} $- Lebesque measure.
Consider the domains $\Omega $, for which the following condition
is fulfilled
\begin{equation}
\mathop {lim}\limits_{t \to + 0} t^{ - \alpha m}v_{M,m} \left( {t}
\right) > 0,\text{ where} \quad \alpha \ge \frac{{n - m}}{{nm}}
\end{equation}

In case of $m = 1$ this condition coincides with classical
isoperimetric conditions. Therefore we condition (5)
 will be call isoperimetric condition.

There is another variant of the Wiener criterion problem , known
among specialists in nonlinear potential theory. $A$ set $\Omega
\subset R^{n}$is said to be $m$-thin at a point if $x_{0} \in
R^{n}$ if $W_{m} \left( {B_{t} \left( {x_{0}}  \right)\backslash
\Omega ,x_{0}}  \right) < + \infty $. This concept of thinness was
first considered in nonlinear potential theory by $\left[ {11}
\right]$. Note that because each sigleton is of $m$-conductuvity
zero it does not have any effect on the $ (\bar B = \bar B(x_0
,r)){\kern 1pt} {\kern 1pt} {\kern 1pt} {\kern 1pt} {\kern 1pt} m
$-thinness of $\Omega $ whether or not the point $x_{0} $ is in
$\Omega $. Also it is trivial $\Omega $ that is $m$ -thin at each
point in the complement of $\overline {\Omega}  $. The sets that
are $m$-thin at $x_{0} $ were characterze as those sets whose
complements are $A$-fine neighborhoods of $x_{0} $. Here $A$-fine
refers to the fine topology of $A$-superharmonic functions.
However it remained unsolved of the $m$ -thinness is equivalent to
the so called Cartan property: `` there is an $A$-superharmonic
function $u$ in neighborhood of $x_{0} $ such that $\mathop
{\mathop {lim}\limits_{x \to x_{0}} } \limits_{x \in \Omega}
infu\left( {x} \right) > u\left( {x_{0}} \right)$.

\bigskip

\textbf{Theorem 1.2.} \textbf{\textit{Let} }$\Omega \subset
R^{n}$\textbf{\textit{ and}}$x_{0} \in {{\overline {\Omega} }
\mathord{\left/ {\vphantom {{\overline {\Omega} }  {\Omega} }}
\right. \kern-\nulldelimiterspace} {\Omega} }$\textbf{\textit{.
Then }}$A$\textbf{\textit{ is} }$m$\textbf{\textit{-thin at}
}$x_{0} $\textbf{\textit{ if and only if there is an
of}}$A$\textbf{\textit{ -superharmonic function}
}$u$\textbf{\textit{ in a neighborhood of}}$x_{0} \in \overline
{\Gamma _{1}}  \cap \Gamma _{2} $\textbf{\textit{ such that} }

 \begin{equation}
 \mathop {\mathop {lim}\limits_{x \to x_{0}} } \limits_{x \in \Omega}
infu\left( {x} \right) > u\left( {x_{0}} \right)
\end{equation}

The proofs of \textbf{Theorems 1.1} and \textbf{1.2} are based on pointwise
estimates of solutions to

\begin{equation}
\label{eq5}
Au = \mu
\end{equation}

\noindent
with a Radon measure $\mu $ on the right side.

The letter $c$ stands for various constants. For an open (closed)
ball $B = B\left( {x_{0},r} \right)$ $\left( {\overline {{\rm B}}
= \overline {B} \left( {x_{0} ,r} \right)} \right)$ with radius
$r$ an center $x_{0} $ and $\sigma > 0$ we write $\sigma {\rm B}$
for the open ball with radius $\sigma \,r$. The barred integral
$sign\,\mathop {\rlap{--} {\smallint} }\limits_{E} fdx$ stands for
the integral average $\left| {E} \right|^{ - 1}\int\limits_{E}
{fdx} $, where $\left| {E} \right|$ is Lebesgue measure of $E$.

The operator $T$ is defined such that for each $\varphi  \in
C_0^\infty  (\partial \Omega \backslash \Gamma _2 )$

\[
Tu(\varphi ) = \int\limits_\Omega  {Au\nabla \varphi dx} ,
\]

\noindent
where $u \in W_{m,0,loc}^{1} \left( {\Omega}  \right)$.
In other words

\[
Tu = - divAu
\]

\noindent
in the sence of distributions.

A solution $u \in W_{m,0,loc}^{1} \left( {\Omega}  \right)$ to the
equation

\begin{equation}
\label{eq6}
Tu = 0
\end{equation}

\noindent always has a continuous representative; we call
continuous solutions $u \in W_{m,0,loc}^{1} \left( {\Omega}
\right) \cap C\left( {\Omega}  \right)$ of (\ref{eq5})
$A$-harmonic in $\Omega $.

A lower semicontinuous function $u:\Omega \to \left( { - \infty ,\infty}
\right]$ is $A$-superharmonic if $u$ is not identically infinite in each
component of $\Omega $, and if for all open $D \subset \subset \Omega $and
$h \in C\left( {\overline {D}}  \right),\,\,\,A$-harmonic in $D,\,\,\,h \le
u\,\,$on $\partial D$ implies $h \le u$ in $D$. A function $v$ is
$A$-subharmonic if -$v$ is $A$-superhamonic.

Clearly, $min\left( {u,v} \right)$and $\lambda u + \sigma $
are$A$-superharmonic if $u$ and $v$ are, and. The following proposition
connects $A$-superharmonic functions with supersolutions of (\ref{eq5}).

\textbf{Proposition 1.} \textbf{\textit{(i) If} }$u \in W_{m,0,loc}^{1}
\left( {\Omega}  \right)$\textbf{\textit{ is such that} }$Tu \ge
0$\textbf{\textit{, then there is an} }$A$\textbf{\textit{-superharmonic
function} }$v$\textbf{\textit{ such that} }$u = v$\textbf{\textit{ a.e.
Moreover,}}
\begin{equation}
v\left( {x} \right) = ess\mathop {lim}\limits_{y \to x} infv\left(
{y} \right)\,\,\,\, \text{for all} \,\,x \in \Omega
\end{equation}

\textbf{\textit{(ii) If} }$v$\textbf{\textit{ is
}}$A$\textbf{\textit{-superharmonic, then (9) holds. Moreover,}
}$Tv \ge 0$ \textbf{\textit{if} }$v \in W_{m,0,loc}^{1} \left(
{\Omega} \right)$\textbf{\textit{ .}}

\textbf{\textit{(iii) If}}$v$\textbf{\textit{ is} }${\rm
A}$\textbf{\textit{-superharmonic and locally bounded, then}}$v \in
W_{m,0,loc}^{1} \left( {\Omega}  \right)$\textbf{\textit{ and} }$Tv \ge
0$\textbf{\textit{.}}

The prove this proposition analogously the prove proposution 2.7
in [10] .

Let $u \in W_{m,0,loc}^{1} \left( {\Omega}  \right)$ be an ${\rm
A}$ -superharmonic function in $\Omega $. Then it follows from
Proposition 1 that $\mu = Tu$ is a nonnegative Radon measure on
$\Omega $. If ${\Omega} '$is an open subset of $\Omega $ with $u
\in W_{m}^{1} \left( {{\Omega} '} \right)$ , the restriction $v$
of $\mu $ to ${\Omega} '$ belongs to the dual space $(
{W_{m,0}^{1} \left( {{\Omega} '} \right))}'$ of $W_{m,0}^{1}
\left( {\Omega}  \right)$. By a standard approximaton we see that

\begin{equation}
\label{eq7}
\int\limits_{\Omega}  {Au\nabla \varphi \,dx} = \int\limits_{{\Omega} '}
{\varphi \,d\mu}
\end{equation}

\noindent for any test function $\varphi \in W_{m,0}^{1} \left(
{{\Omega} '} \right)$, where the last integral is the duality
pairing between $\varphi \in W_{m,0}^{1} \left( {{\Omega} '}
\right)$ and $v \in \left( {W_{m,0}^{1} \left( {{\Omega} '}
\right)}\right)'$.

For the reader's convenience we record here an appropriate form of
weak Harnack inequality (see [9],[12]
 and Proposition 1 above).

Lemma 1.1. Let $B = B\left( {x_{0},z} \right)$ and let $u$ be a
nonnegative $A$- superharmonic function in$3B$. If $q > 0$ is such
that $q\left( {n - p} \right) > n\left( {p - 1} \right)$, then

\[
\left( {\mathop {\rlap{--} {\smallint} }\limits_{2B} u^{q}dx}
\right)^{\frac{{1}}{{q}}} \le c\mathop {inf}\limits_{B} u
\]

\noindent where $c = c\left( {n,m,q}\right) > 0.$

Later we establish estimates for $A$- superharmonic solutions of (\ref{eq5}) in
terms of the Wolff potential

\[
W_{1,m}^{\mu}  \left( {x_{0} ,r} \right) = \int\limits_{o}^{r} {\left(
{\frac{{\mu \left( {B\left( {x_{0} ,t} \right)} \right)}}{{t^{n - m}}}}
\right)} ^{{{1} \mathord{\left/ {\vphantom {{1} {m}}} \right.
\kern-\nulldelimiterspace} {m}} - 1}\frac{{dt}}{{t}}
\]

One easily infers hat $W_{1,2}^{\mu}  \left( {x_{0} ,\infty}
\right)$ is the Newtonian potential of $\mu $. This estimation
gives a solid link between the two nonlinear potential theories.

\textbf{Theorem1.3.} \textbf{\textit{Suppose that} }$u$\textbf{\textit{ is a
nonnegative} }$A$\textbf{\textit{- superharmonic function in}}$B\left(
{x_{0} ,3r} \right)$\textbf{\textit{. If}}$\mu = Tu$\textbf{\textit{, then
}}

\[
c_{1} W_{1,m}^{\mu}  \left( {x_{0} ,r} \right) \le u\left( {x_{0}}  \right)
\le c_{2} infu + c_{3} W_{1,m}^{\mu}  \left( {x_{0} ,2r} \right)
\]

\bigskip

\textbf{\textit{where} }$c_{1} ,c_{2} ,c_{3} $\textbf{\textit{ are
positive constants, depending only on} }$n,m,$\textbf{\textit{ and
the structural constants. In particular,} }$u\left( {x_{0}}
\right) < \infty $\textbf{\textit{ if and only if} }$W_{1,m}^{\mu}
\left( {x_{0} ,r} \right) < \infty$\textbf{.}

Generally speaking is possible indicate that the necessity of the Wiener
test follows from an estimate like that in Theorem1.3. In the present paper
we choose another route, more natural and direct.

Moreover, we deduce from Theorem1.3 a Harnack inequality for positive
solutions to (\ref{eq5}), where the measure $\mu $ satisfies for some positive
constants $ñ$ and $\varepsilon $

\begin{equation}
\label{eq8}
\mu \left( {B\left( {x,r} \right)} \right) \le cr^{n - m + \varepsilon}
\end{equation}

\noindent
whenever $B\left( {x,r} \right)$ is a ball. Iterating the Harnack inequality
in a standard way one sees that the solutions are Holder continuous ;
moreover, we show that if the solutions of $Tu = \mu $ is Holder continuous,
then $\mu $ satisfies a restriction like (\ref{eq7}). As a further consequence of
Theorem1.3 we characterise continuous ${\rm A}$- superharmonic functions in
terms of the corresponding Wolff potentials.

\section{${\rm A}$ -potensials and $m$-conductuvity estimates}

If $r > 0$ and $r \le R$, then there is a positive constant $c_{i}
$ depending only on $n$ and $m$ such that for all $x \in R^{n}$

\[
c^{ - 1}r^{n - m} \le C_{m} \left( {B\left( {x,r} \right),B\left(
{x,R} \right),B\left( {x,r} \right)} \right) \le cr^{n - m}
\]

We say that a conductor $K$ is of $m$-conductuvity zero if

\[
C_{m} \left( {F \cap B,2B,G \cap B} \right) = 0
\]

\noindent
whenever $B$ is an open ball in$R^{n}$. Equivalently $K$
is of $m$-conductuvity zero if and only if

\[
C_{m} \left( {F \cap \Omega ,\Omega ,G \cap \Omega}  \right) = 0
\]

\noindent
for all open sets $\Omega $. Moreover, for $m < n$ this is further
equivalent to

\[
C_{m} \left( {F,R,G} \right) = 0.
\]

We say that a property holds $m$-quasieverywhere in $\Omega $ if
it holds in $\Omega $ except on a set of $m$-conductuvity zero. It
is well known that each function $u \in W_{m,0}^{1} \left(
{\Omega}  \right)$ has a representative for which the limit
$\mathop {lim}\limits_{r \to 0} \mathop {\rlap{--} {\int}
}\limits_{B(x,r)}udy$ exists and equals $u(x)\,\,\,m$
-quasieverywhere in $\Omega$.These representative are called
$m$-refined. In what follows we usually consider only the
$m$-refined representatives of functions in $W_{m,0}^{1} \left(
{\Omega}  \right)$. Note that for a locally bounded ${\rm
A}$-superharmonic function $u$ the limit above exists and is equal
to $u\left( {x} \right)$ for every $x$.

Suppose that $F,G$ be a subset of $\Omega $. For $x \in \Omega $
let

\[
R_{F,G}^{1} \left( {\Omega ,A} \right)\left( {x} \right) = infu\left( {x}
\right)
\]

\noindent where the infimum is taken over all nonnegative ${\rm
A}$-superharmonic functions $u$ in $\Omega $ such that $u \ge 1$
on $Fu = 0$ on $\Omega \backslash G$. The lower semicontinuous
regularization

\[
R_{F,G}^{1} \left( {\Omega ,A} \right)\left( {x} \right) = \mathop
{lim}\limits_{r \to 0} \mathop {inf}\limits_{B_{r}}  R_{F,G}^{1} \left(
{\Omega ,A} \right)
\]

\noindent of $R_{F,G}^{1} \left( {\Omega ,A} \right)$ is called
the $A$-potential of $F$ in $\Omega $. The ${\rm A}$-potential
$\overline {R} _{F,G}^{1} \left( {\Omega ,{\rm A}} \right)$ is
$A$-superharmonic in $\Omega $ and ${\rm A}$-harmonic in $ \Omega
\backslash \bar F$. If $\Omega$ is a bounded and $F,G, \subset
\subset \Omega $, then the ${\rm A}$-potential $u$ of F belongs
to$W_{m,0}^{1} \left( {\Omega}  \right)$ and

\[
C_{m} \left( {F,\Omega ,G} \right) \le \int\limits_{\Omega}
{\left| {\nabla u} \right|} ^{m}dx \le k_{1}^{m} C_{m} \left(
{F,\Omega ,G} \right),
\]

\noindent
(see [13]).

Now we derive estimates for $A$-superharmonic functions in terms
of their Wolff potentials. Because an $A$-superharmonic function
does not necessarily belong to $W^{1}_{m,0,loc} \left( {\Omega}
\right)$, we extend the definition for the operator $T$. If $u$ is
an $A$-superharmonic function in $\Omega $. Then we define

\[
Tu\left( {\varphi}  \right) = \int\limits_{\Omega}  {\mathop {lim}\limits_{k
\to \infty}  A\left( {min\left( {u,k} \right)} \right)\nabla \varphi {\kern
1pt} {\kern 1pt} {\kern 1pt} {\kern 1pt} {\kern 1pt} dx}
\]

\noindent
$\varphi \in W_{m,0}^{1} \left( {\Omega}  \right).$ By
[14] $\mathop {lim}\limits_{k \to \infty} A\left( {min\left( {u,k}
\right)} \right)$is locally integrable and hence $ - Tu$ is its
divergence. Since $min\left( {u,k} \right) \in W^{1}_{m,0,loc}
\left( {\Omega}  \right)$ and $min\left( {u,k} \right) = min\left(
{u,j} \right)$ a.e. in $\left\{ {u < min\left( {k,j} \right)}
\right\}$, the limit exists. It is equal to $A\left( {u} \right)$
if $u \in W_{1,0,loc}^{1} $, which is always the case if $m > 2 -
1\backslash n$.

If $u$ is $A$-superharmonic in $\Omega $, there is nonnegative
Radon measure $\mu $ such that in $\Omega $, and conversely, given
a finite measure $\mu $ in bounded $\Omega $, there is
$A$-superharmonic function $u$ such that $Tu = \mu $ in $\Omega $
and $min\left( {u,k} \right) \in W_{m,0}^{1} \left( {\Omega}
\right)$ for all integers $k$.

We proof auxiliary estimate.

Lemma2.1. Suppose that $u$ is $A$-superharmonic in a ball $B_{2r} \left( {x}
\right)$ and $\mu = Tu$. If $a$ is real constant,$d > 0$ and $m - 1 < \gamma
< n\left( {m - 1} \right)/\left( {n - m + 1} \right)$, then there are
constants $q = q\left( {m,\gamma}  \right)$and

 $c > 0$ such that

\[
\left( {d^{ - r}r^{ - n}\int\limits_{B_{r} \cap \left( {u > a} \right)}
{\left( {u - a} \right)^{r}dx}}  \right)^{m/q} \le cd^{ - r}r^{ -
n}\int\limits_{B_{2r} \cap \left( {u > a} \right)} {\left( {u - a}
\right)^{r}dx} + cd^{1 - m}r^{m - n}\mu \left( {B_{2r}}  \right) \quad ,
\]

\noindent
provided that

\begin{equation}
\label{eq9}
\left| {B_{2r} \cap \left\{ {u > a} \right\}} \right| < \frac{{1}}{{2}}d^{ -
r}\int\limits_{B_{r} \cap \left( {u > a} \right)} {\left( {u - a}
\right)^{r}dx}
\end{equation}

\textbf{\underline {Proof.}} We assume that $u$ is locally bounded and hence
$u \in W_{m,0,loc}^{1} \left( {B_{2r}}  \right)$, without loss of

\noindent
generality that $a = 0$. Let $q = \frac{{m\gamma} }{{m - \gamma /\left( {m -
1} \right)}}$. Notice that $m < q < \frac{{mn}}{{n - m}} = m\ast $.

Using (\ref{eq7}) we obtain

\[
d^{ - r}\int\limits_{B_{r} \cap \left( {0 < u < d} \right)} {u^{r}dx} \le
\left| {B_{r} \cap \left\{ {u > 0} \right\}} \right| \le \left| {B_{2r} \cap
\left\{ {u > 0} \right\}} \right| \le \frac{{1}}{{2}}d^{ -
r}\int\limits_{B_{r} \cap \left( {u > 0} \right)} {u^{r}dx}
\]

\noindent
therefore

\begin{equation}
\label{eq10}
d^{ - r}\int\limits_{B_{r} \cap \left( {u > 0} \right)} {u^{r}dx} \le 2d^{ -
r}\int\limits_{B_{r} \cap \left( {u > d} \right)} {u^{r}dx} \le
c\int\limits_{B_{r}}  {\omega ^{q}dx} \quad ,
\end{equation}

\noindent
where $\omega = \left( {1 + d^{ - 1}u^{ +} } \right)^{r/q} - 1$. Note that
$\nabla \omega = \frac{{\gamma} }{{qd}}\left( {1 + d^{ - 1}u +}
\right)^{r/q - 1}\nabla u^{ +} $.

Let a cut off function $\eta \in C_{0}^{\infty}  \left( {B_{2r}}  \right)$
such that $0 \le \eta \le 1,{\kern 1pt} {\kern 1pt} {\kern 1pt} {\kern 1pt}
{\kern 1pt} {\kern 1pt} {\kern 1pt} {\kern 1pt} {\kern 1pt} {\kern 1pt}
{\kern 1pt} {\kern 1pt} \eta = 1$ on $B_{r} $ and $\left| {\nabla \eta}
\right| \le 2/r$. Using Sobolev inequality we have

\begin{equation}
\label{eq11}
\left( {r^{ - n}\int\limits_{B_{r}}  {\omega ^{q}dx}}  \right)^{m/q} \le
cr^{m - n}\int\limits_{B_{2r}}  {\left| {\nabla \omega}  \right|^{m}\eta
^{m}dx} + cr^{m - n}\int\limits_{B_{2r}}  {{\kern 1pt} \omega ^{m}\left|
{\nabla \eta}  \right|^{m}dx} .
\end{equation}

By substituting the test function $\varphi = \left( {1 - \left( {1 + d^{ -
1}u^{ +} } \right)^{1 - \tau} } \right)u\eta ^{m}$, where $\tau = \gamma
/\left( {m - 1} \right)$,

\noindent
the continuation our estimate, using Young's the quality and (\ref{eq9}) we obtain

\begin{equation}
\label{eq12}
\quad
r^{m}\int\limits_{B_{2r}}  {{\kern 1pt} \omega ^{m}\left| {\nabla \eta}
\right|^{m}dx{\kern 1pt}}  \le c_{3} d^{ - r}\int\limits_{B_{2r} \cap \left(
{u > 0} \right)} {u^{r}dx} \quad .
\end{equation}

Now we remove the assumption that $u$ is locally bounded. For $k > d$ we
write $u_{k} = min\left( {u,k} \right)$ and $\mu _{k} = Tu_{k} $. Then (\ref{eq9})
holds for $u_{k} $ if $k$ is large enough. Hence by the estimates (\ref{eq10})-(\ref{eq12})
we arrive at the estimate

\[
\left( {d^{ - \gamma} r^{ - n}\int\limits_{B_{r} \cap \,\{ u > 0\}}
{u_{k}^{\gamma}  \,dx}}  \right)^{{{m} \mathord{\left/ {\vphantom {{m} {q}}}
\right. \kern-\nulldelimiterspace} {q}}} \le c_{4} d^{ - \gamma} r^{ -
n}\int\limits_{B\,_{2r} \cap \,\{ u > 0\}}  {u_{k}^{\gamma}  \,dx} + c_{4}
d^{1 - m}r^{m - n}\mu _{\,k} \,\,\,\left( {supp\eta}  \right),
\]

\noindent where $c_{4} > 0$. Now letting $k \to \infty $ and using
the weak convergence of $\mu _{\,k} $ to $\mu $ we conclude the
proof.

\textbf{Theorem2.1.} \textbf{\textit{Suppose that} }$u$\textbf{\textit{ is a
nonnegative}}$A$\textbf{\textit{ -superhamonic function in}}$B_{2r} \left(
{x_{0}}  \right)$\textbf{\textit{. If}}$\mu = Tu$\textbf{\textit{, then for
all} }$\gamma > m - 1$\textbf{\textit{ we have that}}

\[
u\left( {x_{0}}  \right) \le c\left( {\mathop {\rlap{--}
{\smallint }}\limits_{B_{r} \left( {x_{0}}  \right)} u^{\gamma}
dx} \right)^{{{1} \mathord{\left/ {\vphantom {{1} {\gamma} }}
\right. \kern-\nulldelimiterspace} {\gamma} }} + cW_{1,m}^{\mu}
\left( {x_{0} ,2r} \right),
\]

\textbf{\textit{where} }$c > 0$\textbf{\textit{ depends at structure.}}

\textbf{Proof.} Let ${{\gamma > n\left( {m - 1} \right)}
\mathord{\left/ {\vphantom {{\gamma > n\left( {m - 1} \right)}
{\left( {n - m + 1} \right)}}} \right. \kern-\nulldelimiterspace}
{\left( {n - m + 1} \right)}}$, fix a constant $\delta \in \left(
{0,1} \right)$ to be a specified later, $B_{l} = B_{r_{\,l}}
\left( {x_{0}}  \right)$, where $r_{j} = 2^{1 - j}r$. We define a
sequence $a_{j}$. Let $a_{0} = 0$ and for $j \ge 0$

\[
a_{j + 1} = a_{l} + \delta ^{ - 1}\left( {r_{l}^{ - n}
\int\limits_{B_{i + 1} \cap \,\{ u > a_{j} \}}  {\left( {u -
a_{j}}  \right)^{\gamma} dx}} \right)^{{{1} \mathord{\left/
{\vphantom {{1} {\gamma} }} \right. \kern-\nulldelimiterspace}
{\gamma} }}
\]

Using Lemma2.1 and accompany estimates we obtain

\[
a_k  - a_1  \le a_{k + 1}  - a_1  = \sum\limits_{j = 1}^k {(a_{j +
1}  - a_1 )}  \le \frac{1}{2}a_k  + c\sum\limits_{j = 1}^k {\left(
{\frac{{\mu \left( {B_j } \right)}}{{r_j^{n - m} }}}
\right)^{1/\left( {m - 1} \right)} }
\]

\noindent
 and hence

\[
\mathop {lim}\limits_{k \to \infty}  a_{k} \le 2a_{1} + c\sum\limits_{j =
1}^{\infty}  {\left( {\frac{{\mu \,\left( {B_{j}}  \right)}}{{r_{j}^{n - m}
}}} \right)} ^{{{1} \mathord{\left/ {\vphantom {{1} {\left( {m - 1}
\right)}}} \right. \kern-\nulldelimiterspace} {\left( {m - 1} \right)}}} \le
c\left( {\mathop {\rlap{--} {\smallint} }\limits_{B_{1}}  u^{\gamma} dx}
\right)^{{{1} \mathord{\left/ {\vphantom {{1} {\gamma} }} \right.
\kern-\nulldelimiterspace} {\gamma} }} + cW_{1,m}^{\mu}  \left( {x_{0} ,2r}
\right)
\]

Now the theorem follows by $infu \le a_{j} \,\,\,\,\,\,\,for\,\,j = 1,2,...$
and for$u$ is lower semicontinuous we conclude that $u\left( {x_{0}}
\right) \le \mathop {lim}\limits_{j \to \infty}  \mathop {inf}\limits_{B_{j}
} u \le \mathop {lim}\limits_{J \to \infty}  infa_{J} $.

\textbf{Proof of Theorem1.3.} The first inequality establishe analogously
$\left[ {10} \right]$. The second inequality follows from Theorem2.1 because
by the weak Harnack inequality in Lemma1.1. We may pick $\gamma \left( {n,m}
\right) > m - 1$ such that

\[
\left( {\mathop {\rlap{--} {\smallint} }\limits_{B_{r}}  u^{\gamma} dx}
\right)^{{{1} \mathord{\left/ {\vphantom {{1} {\gamma} }} \right.
\kern-\nulldelimiterspace} {\gamma} }} \le c\left( {\mathop {\rlap{--}
{\smallint} }\limits_{B_{2r}}  u^{\gamma} dx} \right)^{{{1} \mathord{\left/
{\vphantom {{1} {\gamma} }} \right. \kern-\nulldelimiterspace} {\gamma} }}
\le c\,\mathop {inf}\limits_{B_{r}}  u.
\]

\textbf{Corollary2.1.} \textbf{\textit{Let} } $u$ \textbf{\textit{
be an }} $A$ \textbf{\textit{-superharmonic function in}} $R^{n}$
\textbf{\textit{ with}} $\mathop {inf}\limits_{R^{n}} u =
0$\textbf{\textit{. If}} $\mu = Tu$\textbf{\textit{, then}}

\[
c_{1} W_{1,m}^{\mu}  \left( {x_{0} ;\infty}  \right) \le u\left( {x_{0}}
\right) \le c_{2} W_{1,m}^{\mu}  \left( {x_{0} ;\infty}  \right) \quad ,
\]

\textbf{\textit{where} }$ñ_{1} $\textbf{\textit{ and}}$ñ_{2}
$\textbf{\textit{ are positive constants, depending only on
}}$n,m$\textbf{\textit{ and the structural constants.} }

\textbf{Proof of the Theorem1.2.} The sufficiency part we was
establishe in another paper. We are going to prove the necessity.
Let $K = {{G} \mathord{\left/ {\vphantom {{G} {F}}} \right.
\kern-\nulldelimiterspace} {F}}$ be $m$-thin at $x \notin K$. We
may assume that $K$ is open. Write, $B_{j} = B_{2 - j} \left(
{x_{0}}  \right),\,\,\,r_{j} = 2^{ - j}$, and$K_{j} = K \cap B_{j}
$. Let$\alpha \ge 2$ be an integer, to be specified later. Let $u
= R_{F_{a,G_{a}} } ^{1} \left( {B_{a - 2} :A} \right)$ be the
$A$-potential of $K_{a} $ in $B_{\alpha - 2} $ and$\mu = Tu$. Then
$u \ge 1$ on $K_{\alpha}  $ and it remains to prove that $u\left(
{x_{0}}  \right) < 1$, for some $\alpha $. Using some estimates
$\mu \,\left( {B_{j}}  \right)$ we obtain from Theorem1.3 that

\[
u(x_0 ) \le c{\kern 1pt} {\kern 1pt} \mathop {\inf
}\limits_{B_\alpha  } u + cW_{1,m}^\mu  (x_0 ,r_{\alpha  - 1} )
\le c\sum\limits_{j = \alpha  - 1}^\infty  {\left( {\frac{{C_m
\left( {F_j ,B_{j - 1} ,G_j } \right)}}{{r_j^{n - m} }}}
\right)^{1/(m - 1)} }  \le \frac{1}{2}
\]

Using Theorem1.2 we have that the Cartan property characterizes fine
topologies in nonlinear potential theory. Recall that the $A$-fine topology
is the coarsest topology in$R^{n}$ that makes all $A$-superharmonic
functions in $R^{n}$continuous.

\textbf{Theoem2.2}\textbf{\textit{. Suppose that} } $\Omega
\subset R^{n}$ \textbf{\textit{ and} } $x_{0} \in \overline
{\Omega} $\textbf{\textit{. Then the following are equivalent:
1)}}$x_{0} $ \textbf{\textit{is not an} } $A$\textbf{\textit{-fine
limit point of; 2)}} $\Omega $ \textbf{\textit{ is p-thin}} $x_{0}
$\textbf{\textit{; 3)(Cartan property) There is an} }
$A$\textbf{\textit{-superharmonic function }} $u$
\textbf{\textit{in a neighborhood of} } $x_{0} $ \textbf{\textit{
such that; 4) There are open neighborhood} } $U$ \textbf{\textit{
and }} $V$ \textbf{\textit{ of} } $x_{0} $ \textbf{\textit{ such
that}} $ R_{F \cap U,G \cap U}^1 (V,A) < 1$\textbf{.}

Proof this Theorem follows from Theorem1.2.

Next we are ready to prove Theorem1.1. The notice that the define boundary
regularity we give in $\left[ {14} \right]$.

\textbf{Proof of Theorem1.1.} Suppose that. If $x_{0} $ is an
isolated boundary point, it never is regular as easily follows by
using the maximum principle and the removability of singleton for
bounded $A$-harmonic functions. Hence we are free to assume that
$x_{0} $ is an accumulation point of. Because $E$ is $m$-thin at
$x_{0} $, we now infer from Theorem1.2. that there are balls, such
that and an$A$-superharmonic function $u$ in $B_{2} $ such that,
in and. Next, choose a function such that in and that $\varphi =
1$ in a neighborhood of $x_{0} $. Consider the upper Perron
solution taken in the open set. Because the set of the irregular
boundary points is of conductuvity zero and because it follows
from the generalized comparison principle that in. In particular,.

Hence $x_{0} $ is not regular boundary point of. Since that
barrier characterization for regularity implies that the
regularity is a local property, it follows that is not a regular
boundary point of. Theorem1.1 is proved.

\bigskip

\textbf{Tair Gadjiev}

\textbf{Departament of Nonlinear analysis}

\textbf{Inst.Math.and Mech. NASA}

\bigskip

\textbf{Az.1141 Baku.,st.F.Agaev,9.}

\textbf{E-mail adress; tgadjiev @ mail az.}

\bigskip

\bigskip

\end{document}